\documentstyle[aps,epsfig]{revtex}
\begin{document}
\title{Generating functional for the full parquet approximation}

\author{V.  Jani\v{s}}

\address{Institute of Physics, Academy of Sciences of the Czech Republic,\\
  Na Slovance 2, CZ-18040 Praha 8, Czech Republic\\ e-mail:janis@fzu.cz}

\date{\today}
\maketitle 
\begin{abstract}
  Parquet diagrams sum self-consistently Feynman graphs for the vertex
  function with all two-particle multiple scatterings.  We show how the
  complete parquet equations for the Hubbard-like models can be integrated
  to a generating functional from which all thermodynamic quantities are
  derived via (functional) derivatives.  An explicit Luttinger-Ward
  functional $\Phi[G;\Lambda,{\cal K};U]$ is constructed containing only
  the renormalized one-particle, $G$, irreducible, $\Lambda$, and
  reducible, ${\cal K}$, two-particle propagators as independent
  variational functions.  The parquet approximation is proven to be a
  thermodynamically consistent, $\Phi$-derivable theory obeying the
  necessary conservation laws.
\end{abstract}

One of the major problems in the theory correlated electrons is to
construct in a systematic and controlled way a consistent approximation
interpolating reliably between weak- and strong-coupling regimes.  The two
extreme limits of weak and strong couplings in the archetypal Hubbard model
can be described relatively well. The weak-coupling regime is governed by a
Hartree-Fock mean field with dynamical fluctuations covered by Fermi-liquid
theory. Extended systems at low temperatures are Pauli paramagnets with
smeared out local magnetic moments. For bipartite lattices
antiferromagnetic long-range order sets in at half filling and zero
temperature at arbitrarily small interaction.  In the strong-coupling
regime the Hubbard model at half filling maps onto a Heisenberg
antiferromagnet with pronounced local magnetic moments and the Curie-Weiss
law for the staggered magnetic susceptibility, at least at the mean-field
level. The spectral structure is dominated by separated lower and upper
Hubbard bands and the strongly correlated system seems to be insulating
even in the paramagnetic phase.

However, it is the intermediate coupling, where the effective Coulomb
repulsion is comparable with the kinetic energy and hence neither very weak
nor very strong, that is of great interest for the theorists as well as for
the experimentalists.  At intermediate coupling dynamical fluctuations
control the low-temperature physics of interacting electrons and neither
weak-coupling nor atomic-like perturbation theories are adequate.  In this
nonperturbative regime a singularity in a generic two-particle function is
approached and we expect breakdown of the Fermi-liquid regime and a
transition to an ordered state or eventually to a Mott insulator.

Unfortunately there are only a few techniques for a quantitative analysis
of the transition region between weak and strong coupling. Exact
methods such as the Bethe ansatz or the numerical renormalization group can
be applied only to one-dimensional or single-impurity models
\cite{Lieb68,Wilson75}. Numerical quantum Monte Carlo is good for
thermodynamic properties at relatively high temperatures and is restricted
to small samples \cite{Dagotto94}. Analytic methods are mostly of effective
or mean-field nature \cite{Georges96}. Systematic diagrammatic expansions
usually do not go beyond the fluctuation-exchange approximation (FLEX) that
is known to fail at intermediate coupling \cite{Hamann69}.  Although some
improvements upon or alternatives to the FLEX approximation have been
proposed, the vertex renormalization in these theories remain static
\cite{Bickers91,Vilk94}. The full dynamic and self-consistent vertex
renormalization is contained first in the parquet diagrams
\cite{Dominicis64}.

Parquet diagrams were introduced to describe interaction of mesons more
effectively \cite{Sudakov56}. Since then a number of attempts have been
made to utilize the nontrivial renormalization scheme of the parquet
algebra in condensed matter. Kondo effect \cite{Abrikosov64}, x-ray edge
problem \cite{Roulet69}, formation of the local magnetic moment
\cite{Weiner70} are among the most well known applications. Inability to
solve the parquet equations effectively has impeded broader application of
the method.

Parquet diagrams represent a systematic way of summation and
renormalization of Feynman graphs. Instead of concentrating on the
one-particle irreducible diagrams and the Dyson equation, the parquet
approach sums two-particle diagrams contributing to vertex functions for
which Bethe-Salpeter equations are constructed. The resulting algebra is
much more complicated than that of the one-particle approximations. The
two-particle Green functions obtained from the parquet approximation are
suitable for spectral properties of the system, but it is cumbersome to
gain thermodynamic properties out of them. It has been hitherto unclear
whether the parquet approximation forms a thermodynamically consistent
$\Phi$-derivable approximation fulfilling the necessary conservation laws
\cite{Baym62}.  Only if we construct a generating Luttinger-Ward functional
in closed form from which all the thermodynamic quantities can be derived
via derivatives with respect to auxiliary sources, we can be sure
thermodynamic relations and sum rules are fulfilled.

The author proposed recently a simplification of the full
parquet approximation by summing only two singlet two-particle channels
yielding most divergent diagrams in the critical region of the
metal-insulator transition \cite{Janis98}. A generating functional for such an
approximation was derived via the linked-cluster theorem. The question is
whether also the unabridged parquet equations can be integrated to a
generating thermodynamic potential. It is the aim of this paper to show how
the Luttinger-Ward functional and the grand potential can be constructed
for the complete parquet equations, i.~e. with all two-particle irreducible
diagrams, in lattice models of interacting electrons with a local (Hubbard)
interaction.

There are three topologically inequivalent definitions of a two-particle
irreducibility. It may be defined according to cutting electron-hole or
electron-electron pair propagation, or according to cutting polarization
bubbles shielding the electron-electron interaction. Each possibility
defines a two-particle channel of multiple pair scatterings characterized
by a different binding of independent variables in the vertex functions.  A
general two-particle quantity will be denoted in this paper as in Fig.~1.
Each two-particle function carries three independent four-momenta and two
spin indices. We generally denote the fermionic four-momenta $k=({\bf
  k},i\omega_n)$ and the transferred bosonic ones as $q=({\bf q},i\nu_m)$,
where $\omega_n=(2n+1)\pi T$ and $\nu_m=2m\pi T$ are the respective
Matsubara frequencies at temperature $T=\beta^{-1}$.

It is convenient to introduce a matrix notation in the spin indices to
distinguish different two-particle channels in the parquet diagrams. We
define a $2\times 2$ matrix for the generic two-particle function
$X_{\sigma\sigma'}$:
\begin{eqnarray}
  \label{eq:matrix-def}
  \widehat{X}&=&\left(\begin{array}{cc}
                 X_{\uparrow\uparrow}& X_{\uparrow\downarrow}\\
                  X_{\downarrow\uparrow}&X_{\downarrow\downarrow}
                \end{array}\right)\ . 
\end{eqnarray}
We speak about singlet and triplet contributions to a two-particle function
if the spins of the involved fermions are antiparallel or parallel,
respectively.

We define three matrix multiplication schemes for two-particle quantities
\begin{mathletters} \label{eq:conv-def}
\begin{eqnarray}
  \label{eq:conv-eh}
  \left[\widehat{X}\bullet\widehat{Y}\right]_{\sigma\sigma'}(k,k';q)&=&
  \frac 1{\beta{\cal N}}\sum_{q''} X_{\sigma\sigma'}(k,k';q'')Y_{\sigma
    \sigma'}(k+q'',k'+q'';q-q'')\ ,\\  \label{eq:conv-ee}
  \left[\widehat{X}\circ\widehat{Y}\right]_{\sigma\sigma'}(k,k';q)&=&\frac
  1{\beta{\cal N}}\sum_{q''} X_{\sigma\sigma'}(k,k'+q'';q-q'')Y_{\sigma
    \sigma'}(k+q-q'',k';q'')\ ,\\   \label{eq:conv-U}
  \left[\widehat{X}\star\widehat{Y}\right]_{\sigma\sigma'}(k,k';q)&=&\frac
  1{\beta{\cal N}}\sum_{\sigma''k''}X_{\sigma\sigma''}(k,k'';q)Y_{\sigma''
    \sigma'}(k'',k';q) 
\end{eqnarray}
\end{mathletters}
representing summations in the three inequivalent two-particle channels,
electron-hole, ($eh$), electron-electron, ($ee$), and interaction, ($U$),
respectively. We see that the variables of the two-particle functions are
convoluted differently in inequivalent channels. Note that only the
interaction channel mixes the singlet and triplet contributions.
 
We decompose the full two-particle Green function into a sum of always
reducible and irreducible projections onto each inequivalent channel
\begin{eqnarray}
  \label{eq:K-def}
  {\cal K}_{\sigma\sigma'}(k,k';q)&=&{\cal K}^\alpha_{\sigma
    \sigma'}(k,k';q)+I^\alpha_{\sigma \sigma'}(k,k';q) 
\end{eqnarray}
where $\alpha=eh,ee,U$ refers to a two-particle channel.

The parquet diagrams can at best be represented graphically. Having in mind
the above introduced notation and a general convention that double primed
variables are summed over, we can write
\begin{mathletters} \label{eq:parquet}
  \begin{eqnarray}
   \label{eq:parquet-eh}&&\epsfig{figure=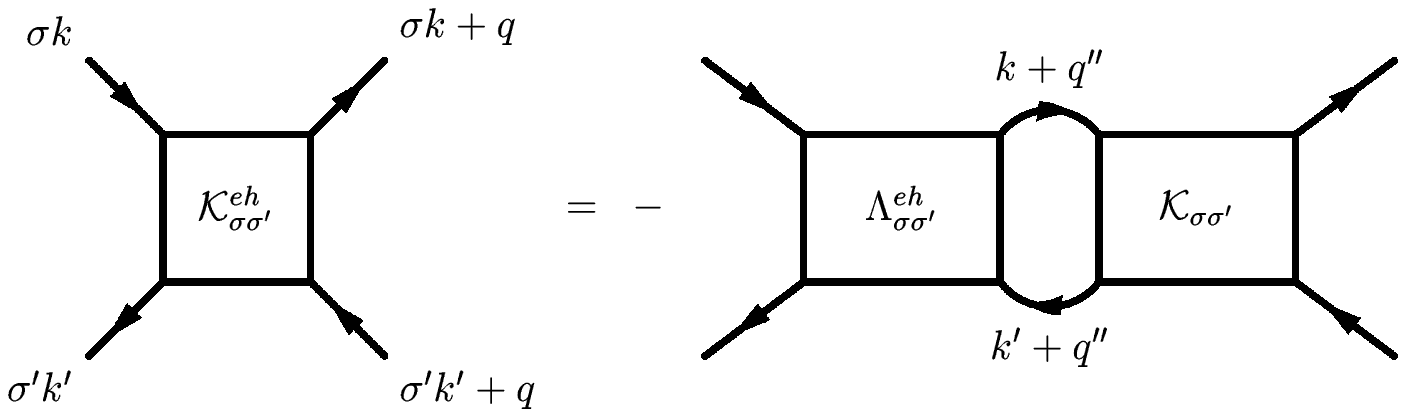,height=4cm}\\ 
   \label{eq:parquet-ee}&&\epsfig{figure=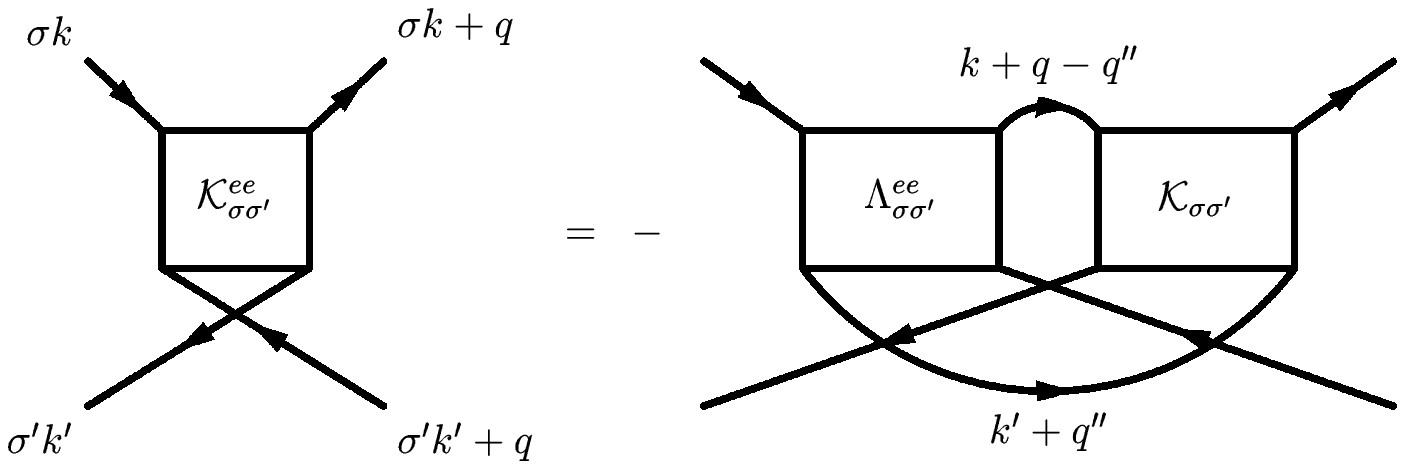,height=4.5cm}\\ 
    \label{eq:parquet-U}&&\epsfig{figure=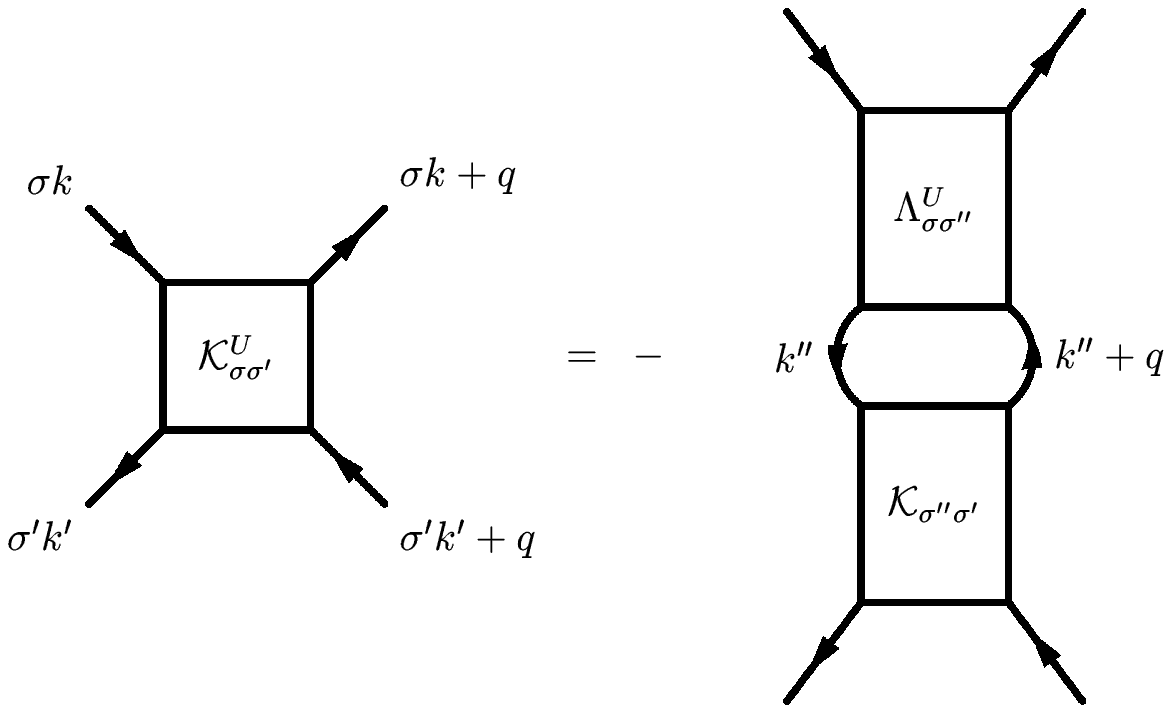,height=6cm} 
  \end{eqnarray}
\end{mathletters}
We labeled only the intermediate four-momenta on the right-hand side of the
parquet equations since the endpoints are the same as those on the
left-hand side.

Equations (\ref{eq:parquet}) are the Bethe-Salpeter equations and
constitute one set of the full parquet algebra. To complete it we must add
relations connecting the two-particle reducible, $\cal K^\alpha$, and
irreducible with two-particle bubbles, $\Lambda^\alpha$, functions. To this
purpose we introduced horizontal ($eh$ and $ee$) and vertical ($U$)
two-particle bubbles
\begin{eqnarray}
  \label{eq:G2}
 G^{(2)h}_{\sigma\sigma'}(k,k';q)=G_\sigma(k+q)G_{\sigma'}(k'+q)\ ,&
 \hspace{1cm}& G^{(2)v}_{\sigma\sigma'}(k,k';q)=G_{\sigma'}(k')G_{\sigma'}
 (k'+q) \ .  
\end{eqnarray}
The  functions $\Lambda^\alpha$ can then be defined as \cite{note1}
\begin{eqnarray}
  \label{eq:Lambda-def}
  \Lambda^\alpha_{\sigma\sigma'}(k,k';q)&=&\left[U\delta_{\sigma',-\sigma}     
   +\Delta I_{\sigma\sigma'}(k,k';q) +\sum_{\alpha''\neq\alpha}{\cal K}
   ^{\alpha''}_{\sigma\sigma'}(k,k';q)\right]G^{(2)\alpha}_{\sigma\sigma'}(k,k';q) 
\end{eqnarray}
where $\Delta I_{\sigma\sigma'}$ is a sum of all Feynman diagrams
simultaneously irreducible in each two-particle channel. The sum contains
only higher-order diagrams where three and more particles are multiply
interconnected. In the usual treatment with only two-particle multiple
scatterings this irreducible part is neglected. We hence put $\Delta
I_{\sigma\sigma'}=0$.
    
The parquet algebra for the two-particle function is now complete.
Equations (\ref{eq:parquet}) and (\ref{eq:Lambda-def}) are to be integrated
to a generating functional. We first construct two functionals being
integrals of equations (\ref{eq:parquet}) and (\ref{eq:Lambda-def}). We can
exclude the reducible parts $\cal K^\alpha$ from (\ref{eq:parquet}) and the
irreducible ones $\Lambda^\alpha$ from (\ref{eq:Lambda-def}). If we denote
the generating functionals for (\ref{eq:parquet}) and (\ref{eq:Lambda-def})
$\Phi_\Lambda$ and $\Phi_K$ respectively, we demand
\begin{eqnarray}
  \label{eq:Phi-dif}
  {\cal K}^\alpha=\frac{\delta\Phi_\Lambda}{\delta\Lambda^{T\alpha}} \ ,
  &\hspace{1cm}& \Lambda^\alpha= \frac{\delta\Phi_K}{\delta{\cal K}^
    {T\alpha}}\ ,
\end{eqnarray}
where $X^{Tv}_{\sigma\sigma'}=X^v_{\sigma'\sigma}$ and
$X^{Th}_{\sigma\sigma'}=X^h_{\sigma\sigma'}$ is a matrix transposition in
the vertical and horizontal channels, respectively.  
The functional $\Phi_\Lambda$ will be a sum of functionals generating each
of the parquet equations in (\ref{eq:parquet}), i.~e. each channel will
have its own generating functional. Recalling the matrix notation and the
multiplication rules (\ref{eq:conv-def}) we can integrate each parquet
equation to the respective generating functional
\begin{mathletters}\label{eq:Phi-int}
\begin{equation}
  \label{eq:Phi-eh}
  \Phi^{eh}_\Lambda=\frac 1{2\beta^2{\cal N}^2}\sum_{\sigma k}\sum_{\sigma'
    k'}\left[\widehat{\Lambda}^{eh}-\frac 12\widehat{\Lambda}^{eh}
    \bullet\widehat{\Lambda}^{eh}-\ln\left(1+\widehat{\Lambda}^{eh}\bullet
    \right)\right]_{\sigma\sigma'} (k,k';0) \ ,
\end{equation}
\begin{equation}
  \label{eq:Phi-ee}
  \Phi^{ee}_\Lambda=\frac 1{2\beta^2{\cal N}^2}\sum_{\sigma k}\sum_{\sigma'
    k'}\left[\widehat{\Lambda}^{ee}-\frac 12\widehat{\Lambda}^{ee}\circ
    \widehat{\Lambda}^{ee} -\ln\left(1+\widehat{\Lambda}^{ee}\circ\right)
  \right]_{\sigma\sigma'}(k,k';0)\ , 
\end{equation}
\begin{equation}
  \label{eq:Phi-U}
  \Phi^{U}_\Lambda=\frac 1{2\beta^2{\cal N}^2}\sum_{\sigma}\sum_{kq}
   \left[\widehat{\Lambda}^{U}-\frac 12\widehat{\Lambda}^{U}\star\widehat
     {\Lambda}^{U} -\ln\left(1+\widehat{\Lambda}^{U}\star\right)\right]_{
     \sigma\sigma}(k,k;q) \ .
\end{equation}\end{mathletters}
The symbols $\widehat{\Lambda}\bullet, \widehat{\Lambda}\circ ,
\widehat{\Lambda}\star$ in the logarithms set the appropriate
multiplication for the power-series definition of the matrix function.

It is much easier to integrate equation (\ref{eq:Lambda-def}).  It is
straightforward to verify that
\begin{eqnarray}
  \label{eq:Phi-K}
  \Phi_K&=&\frac 1{2\beta^2{\cal N}^2}\sum_{\sigma k}\sum_{\sigma'k'}
   \left[U\widehat{\delta}\widehat{G}^{(2)h}\left(\bullet\widehat{\cal
         K}^{eh} +\circ\widehat{\cal K}^{ee}\right)\widehat{G}^{(2)h}+
    \widehat{\cal K}^{eh}\widehat{G}^{(2)h} \bullet  \widehat{\cal K}^{ee} 
     \widehat{G}^{(2)h}\right]_{\sigma\sigma'}(k,k';0)\nonumber\\  
    &&+ \frac 1{2\beta^2{\cal N}^2}\sum_{\sigma}\sum_{kq}\left[\left(U
        \widehat{\delta} +\widehat{{\cal K}}^{eh}+\widehat{{\cal K}}^{ee}
      \right)\widehat{G}^{(2)v}\star\widehat{\cal K}^U\widehat {G}^{(2)v}
    \right]_{\sigma\sigma}(k,k,q),
\end{eqnarray}
where $\widehat{\delta}_{\sigma\sigma'}=\delta_{\sigma',-\sigma}$. Note
that when we multiply functions from the electron-hole and
electron-electron channels the multiplication rule can be taken from either
channel without influencing the result. Connecting the incoming and
outgoing fermion lines in the product is uniquely determined by the charge
conservation.

The Luttinger-Ward functional must finally be stationary with respect to
first variations of the two-particle functions. This can be achieved by
subtracting the mixed $\mbox{tr}\widehat{\Lambda} \widehat{{\cal K}}$ term.
Moreover, the functional must generate the self-energy via a variation with
respect to the full one-electron propagator. We hence must add a purely
one-electron part to the generating functional. The one-electron
contribution is just a second-order term in the free-energy expansion. The
parquet diagrams sum multiple scatterings of pairs of quasiparticles
starting with the next term beyond second order. We choose the correct
overall sign of the generating functional from the self-energy, e.~g. in
second order. The Luttinger-Ward functional for the parquet diagrams
finally reads
\begin{eqnarray}
  \label{eq:Phi-full}
  \Phi[G;\Lambda,{\cal K};U]&=&\frac 1{2\beta^2{\cal
        N}^2}\sum_{\sigma}\sum_{kq}\left[\widehat{\Lambda}^U\star\widehat
      {\cal K}^U  \widehat{G}^{(2)v}\right]_{\sigma \sigma}(k,k,q)\nonumber\\  
    &&\hspace*{-30pt}+\frac 1{2\beta^2{\cal N}^2}\sum_{\sigma k}\sum_
    {\sigma'k'}\left[\left(-\frac 12U\widehat{\delta}\widehat{G}^{(2)h}
        \bullet U\widehat{\delta}+\widehat{\Lambda}^{eh}\bullet\widehat{\cal
          K}^{eh}+\widehat{\Lambda}^{ee}\circ\widehat{\cal K}^{ee}\right)
      \widehat{G}^{(2)h}\right]_{\sigma\sigma'}(k,k'; 0)\nonumber\\[4pt]
&&-\Phi_K\left[G;{\cal K};U\right]-\Phi^{eh}_\Lambda\left[\Lambda^{eh} 
  \right]-\Phi^{ee}_\Lambda\left[\Lambda^{ee}\right]-\Phi^{U}_\Lambda\left[ 
    \Lambda^{U}\right]
\end{eqnarray}
where we introduced in each functional its explicit dependence on the
variational one- and two-particle functions and the bare interaction $U$.

The grand potential generating the complete thermodynamics of the parquet
approximation is constructed from the Luttinger-Ward functional and a
free-electron term with the Hartree contribution \cite{Baym62}. We hence
add also the particle densities as additional variational parameters and
obtain for a fixed $\mu_\sigma=\mu+\sigma h$ where $\mu$ is a chemical
potential and $h$ an external magnetic field
\begin{eqnarray}
\label{eq:Omega}
 \frac 1{{\cal N}}\Omega [n_{\uparrow},n_{\downarrow};\Sigma ,G;\Lambda,
 {\cal K}]&=&-\frac 1{\beta 
   {\cal N}}\sum_{\sigma n,{\bf k}}e^{i\omega_n0^{+}}\Big\{ \ln \left[
     i\omega _n+\mu _\sigma -\epsilon({\bf k}) -Un_{-\sigma }-\Sigma_\sigma
     ({\bf k},i\omega_n)\right] \nonumber\\[4pt]         
& & \hspace*{10pt} +G_\sigma ({\bf k},i\omega_n)\Sigma _\sigma ({\bf
    k},i\omega_n)\Big\} -Un_{\uparrow}n_{\downarrow} +\Phi [G;\Lambda,
{\cal K};U] \ .
\end{eqnarray}
Here $n_{\uparrow},n_{\downarrow};\Sigma ,G;\Lambda,{\cal K}$ are
independent variables and complex functions. Their physical values are
chosen from stationarity of the grand potential (\ref{eq:Omega}) with
respect to variations of its independent variables/functions.  There are
three pairs of Legendre conjugate variational variables in
(\ref{eq:Omega}). The Hartree parameters $n_\uparrow$ and $n_\downarrow$,
the one-electron functions $\Sigma_\sigma(k)$ and $G_\sigma(k)$, and
finally the two-particle vertex functions $\Lambda^\alpha_{\sigma
  \sigma'}(k,k';q)$ and ${\cal K}^\alpha_{\sigma\sigma'}(k,k';q)$.

The stationarity equations for the Hartree parameters lead to a
one-electron sum rule
\begin{eqnarray}
  \label{eq:Hartree}
  n_\sigma&=&\frac 1{\beta {\cal N}}\sum_{n,{\bf k}}\frac{e^{i\omega_n
    0^{+}}}{i\omega _n+\mu _\sigma -\epsilon({\bf k}) -Un_{-\sigma }
    -\Sigma_\sigma ({\bf k},i\omega_n)}\ .
\end{eqnarray}
The stationarity equations for the one-particle functions, i.~e. the
one-electron propagator and the self-energy, measuring in this formulation
corrections to the Hartree approximation, read
\begin{mathletters}
\begin{eqnarray}
  \label{eq:G-equ}
  G_\sigma({\bf k},i\omega_n)&=&\left[i\omega _n+\mu _\sigma-\epsilon
    ({\bf k}) -Un_{-\sigma } -\Sigma_\sigma ({\bf k},i\omega_n)\right]
  ^{-1}\ , \\[4pt]  \label{eq:Sigma}
  \Sigma_\sigma(k)&=&-\frac U{\beta^2{\cal N}^2}\sum_{\sigma'}\sum_{k'q}
  G_{\sigma}(k+q){\cal K}_{\sigma \sigma'}(k+q,k';-q)G_{\sigma'}(k')
  G_{\sigma'}(k'+q) \ .  
\end{eqnarray}\end{mathletters}
Finally, the stationarity equations for the two-particle functions are just
the parquet equations (\ref{eq:parquet}) and (\ref{eq:Lambda-def}) with
$\Delta I_{\sigma\sigma'}=0$.  The full two particle Green function is a
sum of a second-order contribution and the reducible parts from each
two-particle inequivalent channel
\begin{eqnarray}
  \label{eq:K-full}
 {\cal K}_{\sigma\sigma'}(k,k';q)&=&U\delta_{\sigma',-\sigma}
 +{\cal K}^{eh}_{\sigma\sigma'}(k,k';q) +{\cal K}^{ee}_{\sigma\sigma'}
 (k,k';q) +{\cal K}^{U}_{\sigma\sigma'}(k,k';q) \ .
\end{eqnarray}

Thermodynamics of the parquet approximation is completely determined by
(\ref{eq:Phi-int})-(\ref{eq:K-full}) into which appropriate auxiliary
external sources are introduced. All thermodynamic relations are hence
fulfilled in the parquet summation scheme. Moreover due to the full
self-consistency in the one- and two-particle propagators, the parquet
approximation obeys momentum and energy conservation laws \cite{Baym62}.
However, it is unclear whether the parquet approximation conserves also
charge as the only generator of the electrostatic energy. The electrostatic
energy (due to the Coulomb interaction $U$) of the solutions must entirely
be generated by the actual distribution of charge carriers (electron
density $n$).  This conservation law induces a Ward-like identity connecting a
variation in the electrostatic energy with a variation in the mass
distribution \cite{Janis98}. The identity relates a thermodynamic
correlation function defined as a derivative of the one-electron density
with respect to an auxiliary field (electro-chemical potential) and the
full two-particle Green function ${\cal K}_{\uparrow\downarrow}$ from
(\ref{eq:K-full}).
 
The correlation function describing the variation of the electrostatic
energy can be derived from the generating functional if we introduce a
small perturbation of the local Coulomb interaction $U\rightarrow U+\delta
U(q)$.  We define the correlation function as
\begin{eqnarray}
  \label{eq:C-correl}
  {\cal C}(q)&=&\frac{\delta \Phi \left[G;\Lambda,{\cal K};U\right] }{
    \delta U(q)}\ \bigg|_{\delta U=0}\nonumber\\
 &=&-\frac 1{2\beta^2{\cal N}^2}\sum_{kk'}G^{(2)}_{\uparrow
     \downarrow}(k,k';q)\left\{{\cal K}_{\uparrow  \downarrow}(k,k';-q)+
     {\cal K}_{\uparrow\downarrow}(k,k';q) \right\}G^{(2)}_{\uparrow
     \downarrow}(k,k';-q)\ . 
\end{eqnarray}
The static part of this dynamic correlation function is important for the
thermodynamic stability of a particular solution (phase). This function
must be negative in a stable phase
\begin{eqnarray}\label{eq:stability}
   {\cal C}({\bf q},0)&<&0  
\end{eqnarray}
as a consequence of a repulsive character of the Coulomb interaction
\cite{Janis98}.  The correlation function ${\cal C}({\bf q},0)$ is from
definition connected with the magnetic susceptibility $\chi({\bf q})$ and
the compressibility $\kappa({\bf q})$
\begin{eqnarray}
  \label{eq:C-Ward}
  {\cal C}({\bf q},0)&=&\frac 1{\cal N}\sum_{\bf ij}e^{-i({\bf R}_{\bf i}- 
   {\bf R}_{\bf j}){\bf q}}\left[\langle n_{{\bf i}\uparrow}  n_{{\bf j}
       \downarrow}\rangle- \langle n_{{\bf i}\uparrow}\rangle\langle
       n_{{\bf j}\downarrow}\rangle\right]\equiv \frac T4\left[ \kappa
       ({\bf q})-\chi ({\bf q})\right] \ ,
\end{eqnarray}
which together with (\ref{eq:stability}) indicates that in tight-binding
models with a local electrostatic interaction the spin fluctuations are
stronger than the charge ones.
 
Equality (\ref{eq:C-Ward}) is rigorously fulfilled in an exact solution.
The Ward identity due to conservation of sources of the electrostatic
energy demands that sums of both sides of (\ref{eq:C-Ward}) over momenta
must equal, otherwise there are spurious sources of the electrostatic
potential in the system.  Approximate theories with only one-electron
renormalizations break equality (\ref{eq:C-Ward}) qualitatively, i.~e. the
left- and right-hand sides lead to incompatible phase diagrams and it is
unclear how to define two-particle functions consistently. The only way to
improve upon this discrepancy is to introduce dynamical vertex
(two-particle) renormalizations.  It is clear that the exact equality in
(\ref{eq:C-Ward}) can be achieved in self-consistent theories only in a
complete solution where arbitrarily large clusters of particles are
involved. In order to produce unambiguous results for two-particle
correlation functions it must be required from approximate theories that
the left- and right-hand sides of (\ref{eq:C-Ward}) generate the same
divergences and lead to equivalent phase diagrams.

Although the parquet diagrams contain an advanced dynamical renormalization
of the vertex functions, skeleton three-particle diagrams are not involved.
Higher-order skeleton diagrams are generated by derivatives of the
self-energy (\ref{eq:Sigma}) with respect to the electro-chemical potential
when constructing the susceptibility or the compressibility. However,
skeleton diagrams beyond the parquet approximation are fully renormalized
in the two-particle sector, i.~e. they use appropriate projections of the
full two-particle Green function ${\cal K}_{\sigma\sigma'}$ instead of the
bare interaction $U$.  The lowest-order contribution
to $\Delta I_{\sigma\sigma'}$ is shown in Fig.~2. 
Unlike other simpler approximations the two-particle functions in the
parquet approach can show only {\em integrable singularities}. A
singularity can arise in one or more two-particle channels simultaneously.
Only one bosonic variable, the conserved transfer four-momentum in the
multiple scattering events, is relevant in each channel.  It is for the
electron-hole, electron-electron, and interaction channel $k-k'$, $k+k'+q$,
and $q$, respectively. The divergences in two-particle functions are
smeared out (regularized) whenever one integrates over the variables in
which the singularity arises. We realize that three-particle skeleton
diagrams as in Fig.~2 contain integrals over all relevant two-particle
four-momenta simultaneously and are hence finite. Unless {\em qualitatively
  new} singularities arise in three-particle (or higher-order) skeleton
diagrams, the parquet approximation is stable with respect to fluctuations
and the left- and right-hand sides of (\ref{eq:C-Ward}) lead to the same
instabilities and hence to qualitatively the same phase diagram.
Appearance of new singularities in three-particle and higher-order skeleton
diagrams, not seen at the two-particle level, would indicate in the
renormalization-group language that higher-order effective interactions are
relevant. Neither the Bethe ansatz nor the renormalization group suggest
relevance of interactions beyond the two-particle ones. 

Parquet diagrams are a systematic approximation summing all two-particle
irreducible Feynman graphs with fully renormalized one-electron
propagators. The idea behind this summation scheme is to replace an
expansion in the bare interaction strength $U$ by skeleton diagrams with
renormalized two-particle functions. This is a realization of the
field-theoretic renormalization of perturbation theory close to critical
points.  Parquet approximation is the simplest theory with two explicit
vertex functions containing minimally one-loop renormalization of the
critical behavior.  However, one cannot reduce the singular two-particle
Green function to a static running coupling constant as in classical phase
transitions.  As well as we cannot reduce renormalized perturbation theory
to a weak-coupling loop expansion. Due to a complicated structure of
dynamical fluctuations at quantum phase transitions in itinerant systems,
one has to keep dynamical two-particle Green functions in the parquet
equations and their extensions.  Parquet diagrams offer a comprehensive
basis for a reliable assessment of the critical behavior at quantum phase
transitions with singularities in two-particle Green functions.

Extensions of the parquet approximation contain only skeleton diagrams
contributing to $\Delta I_{\sigma\sigma'}$ free of one- and two-particle
insertions.  The interaction strength is everywhere replaced by the
two-particle Green function. The parquet approximation offers an overall
reliable solution whenever the {\em norm} of the full two-particle function
$\parallel{\cal K}G^{(2)}\parallel\ll 1$.  Since the parquet equations allow only
for integrable singularities in the two-particle Green functions, their
solution may obey the reliability criterion even at intermediate and strong
coupling. At least $\parallel{\cal K}G^{(2)}\parallel< 1$ is guaranteed in the
parquet approximation. This feature makes it an appropriate tool for
studying the transition region between the weak-coupling, Fermi-liquid
solution and the strong-coupling, local-moment phase in the lattice models
with an effective local (Hubbard) interaction.

To conclude, we integrated the parquet diagrams for two-particle Green
functions to a generating Luttinger-Ward functional. We demonstrated that
the parquet summation scheme belongs to thermodynamically consistent and
conserving $\Phi$-derivable approximations in Baym's sense if the two
particle functions are generated from appropriate functional derivatives of
the self-energy.  The difference between the thermodynamic (variational
derivatives) and diagrammatic definitions of two-particle functions
(Aslamazov-Larkin diagrams \cite{Bickers89}), reflecting deviations from
exact identity (\ref{eq:C-Ward}), are expected to vanish in leading
asymptotic order at two-particle critical points.

Although the explicit grand potential does not help to resolve the
complicated algebra of dynamical variables coupled in the parquet
equations, it enables one to control systematic expansions beyond and
eventual simplifications within the parquet approximation. The parquet
diagrams and their derivatives have been presented in a new light from
which the role, relevance, and systematics of the parquet-type graphs have
become more transparent.
 
The work was supported by the grant No. 202/98/1290 of the Grant Agency of
the Czech Republic.


\newpage
\begin{figure}
  \hspace*{80pt} \epsfig{figure=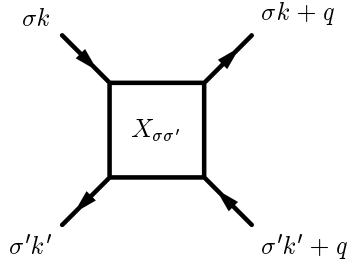,height=4cm}
\caption{Generic two-particle function with three independent
  four-momenta and a defined order of incoming and outgoing fermions.}
\end{figure}
\begin{figure}
  \epsfig{figure=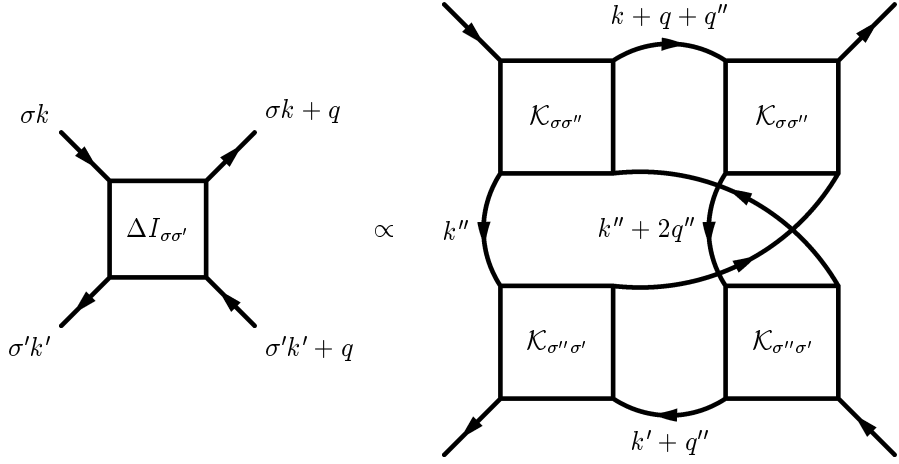,height=6.5cm} \vskip10pt
\caption{Lowest-order contribution not included in the parquet
  approximation. The bare interaction is replaced by the full two-particle
  function from the parquet approximation. Double primed variables are
  summation indices.}
\end{figure} 

\begin{thebibliography}{99}
\bibitem{Lieb68} E.~H.~Lieb and F.~Y.~Wu, \newblock Phys. Rev. Lett. {\bf
    20}, 1445 (1968).
  
\bibitem{Wilson75} K.~G.~Wilson, \newblock Rev. Mod. Phys. {\bf 47}, 773
  (1975).
  
\bibitem{Dagotto94} E.~Dagotto, \newblock Rev. Mod. Phys. {\bf 66}, 763
  (1994).
  
\bibitem{Georges96} A.~Georges, G.~Kotliar, W.~Krauth, and M.~Rozenberg,
  \newblock Rev. Mod. Phys. {\bf 68}, 13 (1996).
  
\bibitem{Hamann69} D.~R.~Hamann, \newblock Phys. Rev. B{\bf 186}, 549
  (1969).
  
\bibitem{Bickers91} N.~E.~Bickers and S.~R.~White, \newblock Phys. Rev.
  B{\bf 43}, 8044 (1991).
  
\bibitem{Vilk94} Y.~M.~Vilk, L.~Cheng, and A.~M.~S.~Tremblay, \newblock
  Phys. Rev. B{\bf 49}, 13~267 (1994).
  
\bibitem{Dominicis64} C.~De~Dominicis and P.~C.~Martin, \newblock J. Math.
  Phys. {\bf 5}, 14 (1964).
  
\bibitem{Sudakov56} V.~V.~Sudakov, \newblock Dokl. Akad. Nauk SSSR {\bf
    111}, 338 (1956), [English transl. Soviet Phys. - Doklady {\bf 1}, 662
  (1957)].
  
\bibitem{Abrikosov64} A.~A.~Abrikosov, \newblock Physics {\bf 2}, 5 (1964).
  
\bibitem{Roulet69} B.~Roulet, J.~Gavoret, and P. Nozi\`eres, \newblock
  Phys. Rev. {\bf 178}, 1078 (1969).
  
\bibitem{Weiner70} R.~A. Weiner, \newblock Phys. Rev. Lett. {\bf 24}, 1071
  (1970).
  
\bibitem{Baym62} G.~Baym, \newblock Phys. Rev. {\bf 127}, 1391 (1962).
  
\bibitem{Janis98} V.~Jani\v{s}, \newblock J. Phys.: Condens. Matter {\bf
    10}, 2915 (1998).
  
\bibitem{note1} Usually the two-particle irreducible functions are defined
  without the one-particle propagators. For the construction of the
  generating functional it is more convenient to absorb the two-particle
  bubble into the bare interaction.

\bibitem{Bickers89} N.~E.~Bickers and D.~J.~Scalapino, \newblock Annals of
  Physics \textbf{193}, 206 (1989)
\end{thebibliography}
\end{document}